\documentclass[12pt]{article}
\usepackage{a4wide}
\usepackage{psfrag}
\usepackage{graphics}
\usepackage{amsfonts}
\usepackage{amstext}
\usepackage{amsmath}
\usepackage{amssymb}
\usepackage{amsthm}
\usepackage[mathscr]{eucal}
\makeatletter
 
 \@addtoreset{equation}{section}
\makeatother     
\newtheorem{theorem}{Theorem}[section]

\def\a{\alpha}
\def\A{\mathcal{A}}
\def\B{\mathcal{B}}
\def\C{\mathcal{C}}
\def\Ai{{\rm Ai}}
\def\b{\beta}
\def\da{\dagger}

\def\g{\gamma}

\def\K{\mathcal{K}}

\def\L{\Lambda}

\def\o{\omega}
\def\O{\Omega}

\def\P{\mathbb{P}}

\def\t{\tau}

\def\Z{\mathbb{Z}}

\begin{document}
\title{\bf Correlation function of the Schur process
with a fixed final partition}
\date{}
\author{
\vspace{5mm}
T. Imamura
{\footnote {\tt e-mail: timamura@iis.u-tokyo.ac.jp}}
~~and T. Sasamoto 
{\footnote {\tt e-mail: sasamoto@math.s.chiba-u.ac.jp}}
\\
{\it $^*$Institute of Industrial Science, University of Tokyo,}\\
\vspace{5mm}
{\it Komaba 4-6-1, Meguro-ku, Tokyo 153-8505, Japan}\\
{\it $^{\dag}$Department of Mathematics and Informatics,}\\
{\it Chiba University,}\\
{\it Yayoi-cho 1-33, Inage-ku, Chiba 263-8522, Japan}\\
}

\maketitle

\begin{abstract}
We consider a generalization of the Schur process 
in which a partition evolves from the empty partition 
into an arbitrary fixed final partition. We obtain a double 
integral representation of the correlation kernel. For a 
special final partition with only one row, the edge scaling 
limit is also discussed by the use of the saddle point analysis. 
If we appropriately scale the length of the row, the limiting 
correlation kernel changes from the extended Airy kernel.
\end{abstract}

\newpage
\section{Introduction}
Recently in nonequilibrium stochastic models which belong to 
the one-dimensional Kardar-Parisi-Zhang(KPZ) universality 
class~\cite{KaPaZh1986}, remarkable progress has been made 
on the understanding of the scaling function. The exact form 
of the function is obtained and its relation to the random matrix 
theory is revealed~\cite{Jo2005L,Sp2006, FePr2006, Ma2006, Sa2007}. 
The common mathematical structure among many of these models 
is that certain ``correlation functions" can be expressed as a 
determinant of a correlation kernel. This structure makes it 
possible to exactly analyze the function. The process having the 
structure is called determinantal process. 

The Schur process, which was introduced in~\cite{OkRe2003}, 
is the typical process of the determinantal processes. 
Let $\lambda$ be a 
partition, i.e. $\lambda=(\lambda_1,\lambda_2,\cdots)$ where
$\lambda_i\in\{0,1,2,\cdots\}$ and $\lambda_1\ge\lambda_2\ge\cdots$.
For a set of partitions $\{\lambda^{(1)},\lambda^{(2)},\cdots,
\lambda^{(4N-1)}\}$, the measure is defined as
\begin{equation}
s_{\lambda^{(1)}}(a^{(1)})
\left(\prod_{j=1}^{2N-1} 
s_{\lambda^{(2j-1)}/\lambda^{(2j)}}(a^{(2j)})
s_{\lambda^{(2j+1)}/\lambda^{(2j)}}
(a^{(2j+1)})\right)
s_{\lambda^{(4N-1)}}(a^{(4N)}),
\label{schurp}
\end{equation}
where $s_{\lambda/\mu}(a^{(i)})$ is the skew Schur function of 
shape $\lambda/\mu$ in the variables $a^{(i)}=(a^{(i)}_1,
a^{(i)}_2,\cdots)$. (For the definition of the Schur function,
see~\eqref{JaTr}.) 
Like the equation above, probability measures described by the Schur 
function appear in various fields in statistical mechanics. It has been 
known that the Schur function is useful for studying the problems of 
many-body nonintersecting random walk and Brownian motion due to 
its determinant structure and the relation to the combinatorics of 
the Young diagram~\cite{GuOwVi1998, Ba2000, Fo2000, NaFo2002,
KaTa2002,KaTa2005}.
Furthermore the weight~\eqref{schurp} and 
its variant appear in analyses of various nonequilibrium processes 
such as the polynuclear growth(PNG) model~\cite{Jo2003}, melting in 
three-dimensional crystal~\cite{FeSp2003}, random tiling 
process~\cite{Jo2005}, 
asymmetric exclusion process~\cite{ImSa2007} and so on. 

In order to visualize this model, we explain many-body random walk 
interpretation of
the process introduced in~\cite{OkRe2003}.
At $t=0$ and $t=4N$, the random walker labeled $i~(i=1,2,\cdots)$ 
is on the position $1-i$ on one dimensional lattice 
$\Z=\{\cdots,-2,-1,0,1,2,\cdots\}$.
We regard $\lambda^{(n)}_i-i+1$ as the position of the $i$th walker, 
at time $n$. Here $\lambda^{(n)}_i$ is the $i$th element of the 
partition $\lambda^{(n)}=(\lambda^{(n)}_1,\lambda^{(n)}_2,\cdots,0,0,\cdots)$.
On the other hand, the Schur function $s_{\lambda^{(n+1)}/\lambda^{(n)}}(a^{(n)})$ 
is interpreted as the transition weight of the 
walkers from the position $\{\lambda^{(n)}_i-i+1\}_{i=1,2,\cdots}$
at time $n$ to $\{\lambda^{(n+1)}_i-i+1\}_{i=1,2,\cdots}$ at time 
$n+1$.  Note that the variables $a^{(n)}$ characterize the transition 
weight at each time step.  

Due to the property $\lambda^{(n)}_1>\lambda^{(n)}_2-1>\cdots>\lambda^{(n)}_i-i+1>
\lambda^{(n)}_{i+1}-i>\cdots$, the walkers do not intersect. Note that
the transition weight (the Schur function) has the Slater determinant 
like structure \eqref{JaTr}. Thus, this nonintersecting random walk
has the free fermionic feature and we find that \eqref{schurp}  
describes the (free fermionic) model of nonintersecting random 
walkers which depart from the positions $(0,-1,-2,\cdots)$ at time 
$0$ and arrive at the same positions at time $4N$ by way of the 
positions $\{\lambda^{(n)}_i-i+1\}_{i=1,2,\cdots}$ at time 
$n=1,2,\cdots,4N-1,$ as depicted in Fig.~1. 

This type of nonintersecting random walk is closely related to the
PNG model. In~\cite{Jo2003,PrSp2002b}, the authors introduced the 
same type of the random walk model called the multilayer PNG model 
in order to discuss the equal-time multipoint correlations of height 
fluctuations. In this model, the position of the first walker 
$\lambda_1^{(n)}$ at time $n,~(n=1,\cdots,4N)$
represents the height at position $n$ and at some fixed time. 
Note that in order to find the dynamics of the first walker,
we need the information about the correlation among the first walker and
the other walkers. Thus, we focus on the correlation function (see~\eqref{cor})
of the Schur process in this paper. We are also interested in
seeing how the initial or final configurations of the walkers 
affect the dynamical behavior.
In the study of the PNG model with an 
external source~\cite{BaRa2000, ImSa2004, ImSa2005},
it has been known that the property of the height fluctuation 
changes if the external source is larger than a critical value.
As we discussed in~\cite{ImSa2005}, the external source in the PNG model
may be related to the initial or final configurations of 
the nonintersecting random walk model.  

In order to discuss the problem, 
we investigate the following process in this paper: 
\begin{equation}
s_{\lambda^{(1)}}(a^{(1)})
\left(\prod_{j=1}^{2N-1}
s_{\lambda^{(2j-1)}/\lambda^{(2j)}}(a^{(2j)})
s_{\lambda^{(2j+1)}/\lambda^{(2j)}}(a^{(2j+1)})\right)
s_{\lambda^{(4N-1)}/\mu^{(4N)}}(a^{(4N)}).
\label{pschurp}
\end{equation}
Here the partition $\mu^{(4N)}=(\mu^{(4N)}_1,\mu^{(4N)}_2,\cdots,
\mu^{(4N)}_n,0,0,\cdots)$ is arbitrary but fixed. Notice that
the weight~\eqref{pschurp} is reduced to~\eqref{schurp} if we choose
$\mu^{(4N)}=\phi:=(0,0,\cdots)$. In the interpretation of 
nonintersecting random walk, the partition $\mu^{(4N)}$ corresponds 
to the final 
configuration $\{\mu^{(4N)}_i-i+1\}_{i=1,2,\cdots}$. 
Thus, the weight~\eqref{pschurp} is a natural extension of the 
Schur process. Fig.~2 illustrates the situation. 
In this paper, we obtain the double integral formula 
of the correlation kernel of the (dynamical) correlation function 
(see~\eqref{rcor}). Furthermore, 
by the use of the formula, we discuss the asymptotic limit of the 
correlation function. It will be shown that if we appropriately scale 
the partition $\mu^{(4N)}$, the scaling limit changes from 
case~\eqref{schurp}.

In~\cite{OkRe2003}, Okounkov and Reshetikhin originally derived
determinant structure of the correlation function and  
the integral representation of the correlation kernel in the 
Schur process~\eqref{schurp} by the method of using the fermion 
operators.  This method was first developed in their previous 
article~\cite{Ok1999}.
In~\cite{Jo2003}, Johansson considered the height fluctuation property
of the (multilayer) PNG model which is essentially the same as  
process~\eqref{schurp} and discussed the derivation of the integral
formula by using a property of a Toeplitz matrix.
Furthermore, the edge scaling limit(see Sec. 2.3) of the 
correlation kernel is also obtained.
A simple linear algebraic proof of the integral formula along 
the approach in~\cite{Jo2003} and its Pfaffian analog were 
discussed by Borodin and Rains~\cite{BoRa2006}.
In the noncolliding Brownian motion, which we can
regard as the continuum version of Fig.1, 
the correlation function and its asymptotic form have been discussed 
in the similar situation to Fig.~1 where Brownian particles start 
and end at one point. It is found that the process is closely 
related to the random matrix theory and
orthogonal polynomials~\cite{KaTa2002, KaTa2004, KaTa2007p}.
The connection between the dynamics of the first particle and
Painlev{\'e} equation is also revealed~\cite{TrWi2004, TrWi2007}.

In case~\eqref{pschurp}, on the other hand, the asymptotics of the 
correlation function has not been discussed yet since an integral 
representation of the correlation kernel has not been obtained. 
In this paper, we discuss the generalization of the method
in~\cite{Jo2003} in order to obtain the integral formula 
for the correlation kernel in process~\eqref{pschurp}
although we could consider the generalization of the methods 
in~\cite{OkRe2003, BoRa2006}. 
In the similar situation of noncolliding Brownian motion 
where the particles start from one initial point and converge 
to fixed final points at the end, it is revealed that the process 
is related to the random matrix model with external
source and due to the properties of multiple orthogonal 
polynomials, its asymptotic limit can also be 
obtained~\cite{ApBlKu2005, ImSa2005}.
Recently a partial differential equation describing the dynamics 
of the first particle is also obtained in this situation~\cite{AdDeMo2007p}.
However, process~\eqref{pschurp} includes the parameters 
$a^{(i)}$ by which we can change the transition weight every time 
step. Therefore, \eqref{pschurp} is a more general process in the 
sense that it includes both $a^{(i)}$ describing 
the temporary inhomogeneity and boundary parameter $\mu^{(4N)}$. 

This paper is arranged as follows.
In Sec.~\ref{sec2}, we discuss the background of this study especially
the result about the process defined by products of 
determinant~\cite{Jo2003} and give our main results: the double integral
formula of  the correlation kernel (Theorem~\ref{th1}) and its edge scaling 
limit (Theorem~\ref{th2}). 
The double integral formula of the correlation kernel is derived in Sec.~\ref{proof1}.
In Sec.~\ref{asan}, we discuss the edge scaling limit of the correlation kernel
applying the saddle point method to the integral formula.
The Concluding remark is given in Sec.~\ref{con}. 

\section{Correlation function}\label{sec2}
\subsection{Determinant representation of correlation function}
The determinantal structure of correlation function 
has been derived for various stochastic processes such as the Schur 
process~\eqref{schurp}~\cite{OkRe2003} and random 
matrices~\cite{EyMe1998}.  
For later discussion, we consider a class of general measures
which contains the above measures and describe 
the result obtained in~\cite{Jo2003}. 
This is obtained
by generalizing the discussion about the random matrix theory 
in~\cite{TrWi1998}.

For the set $\{x^{(i)}\}_{i=1,\cdots,4N-1}$, where
$x^{(i)}=(x^{(i)}_1,x^{(i)}_2,\cdots, x^{(i)}_{M})$,
we consider the weight defined by the products of determinants,
\begin{equation}
w_{M,4N}(\{x^{(i)}\}_{i=1,\cdots,4N-1})
=\prod_{r=0}^{4N-1} \det\left(\phi_{r,r+1}(x^{(r)}_i,x^{(r+1)}_j)
\right)_{i,j=1}^M,
\label{pdet}
\end{equation}
where $\{\phi_{r,r+1}(x,y)\}_{r=1,\cdots,4N}$ are some functions
on $\Z^2$ and
we fix $x^{(0)}$ and $x^{(4N)}$
such that
\begin{equation}
x^{(k)}_1>x^{(k)}_2>\cdots>x^{(k)}_M,~~k=0,4N.
\label{0cond}
\end{equation} 
(In~\cite{Jo2003}, some condition is 
assumed for the function $\phi_{r,r+1}(x,y)$ such that all 
objects that appear in the discussion converge.) 
 
By using Lindstr{\"o}m-Gessel-Viennot's method~\cite{Li1973,GeVi1989}
(see also~\cite{KaMc1959}), we find that under
condition~\eqref{0cond},  
the weights assigned for all configurations but those
satisfying
\begin{equation}
x^{(i)}_1>x^{(i)}_2>\cdots>x^{(i)}_M
\label{xcondition}
\end{equation}
vanish. Namely, when we interpret $x^{(j)}_i$ as the position 
of the particle labeled $i$ at time $j$ as in the previous
section, the weight does not vanish only in the case
where the particles do not intersect.

The correlation function of the measure~\eqref{pdet}
$R(x^{(1)}_1,\cdots,x^{(1)}_{k_1},\cdots,x^{(4N-1)}_1,\cdots,
x^{(4N-1)}_{k_{4N-1}})$ is defined as
\begin{multline}
R(x^{(1)}_1,\cdots,x^{(1)}_{k_1},\cdots,x^{(4N-1)}_1,\cdots,
x^{(4N-1)}_{k_{4N-1}})\\
=\frac{1}{Z}\left(\prod_{i=1}^{4N-1}\sum_{x^{(i)}_{k_i+1},\cdots
,x^{(i)}_{M}=-\infty}^{\infty}\right)
w_{M,4N}(\{x^{(i)}\}_{i=1,\cdots,4N-1}),
\label{cor}
\end{multline}
where $Z=\prod_{i=1}^{4N-1}\sum_{x^{(i)}_{1},\cdots
,x^{(i)}_{M}=-\infty}^{\infty} w_{M,4N}(\{x^{(i)}\}_{i=1,\cdots, 4N-1})$ is 
the normalization constant. In~\cite{Jo2003}, Johansson showed 
that it can be represented as the determinant,
\begin{multline}
R(x^{(1)}_1,\cdots,x^{(1)}_{k_1},\cdots,x^{(4N-1)}_1,\cdots,
x^{(4N-1)}_{k_{4N-1}})\\
=
\det\left(
K(r,x^{(r)}_{j_r};s,x^{(s)}_{j_s})
\right)_{1\le r,s\le 4N-1, 0\le j_r\le k_r, 
0\le j_s\le k_s}.
\label{rcor}
\end{multline}
The correlation kernel $K(r_1,x_{1}; r_2,x_{2})$ is expressed as
\begin{align}
K(r_1,x_1;r_2,x_2)=\tilde{K}(r_1,x_1;r_2,x_2)
-\phi_{r_1,r_2}(x_1,x_2).
\label{kernel}
\end{align}
Here,
\begin{align}
&\phi_{r,s}(x,y)=
\begin{cases}
\sum_{x_{r+1}=-\infty}^{\infty}\cdots
\sum_{x_{s-1}=-\infty}^{\infty} \phi_{r,r+1}(x,x_{r+1})\cdots
\phi_{s-1,s}(x_{s-1},y),& \text{for~}r<s,\\
0,&\text{for~}r\ge s,
\end{cases}
\label{dphi}\\
&\tilde{K}(r_1,x_1;r_2,x_2)
=\sum_{i,j=1}^{M}\phi_{r_1,4N}(x_1,x_i^{(4N)})
(\A^{-1})_{i,j}
\phi_{0,r_2}(x^{(0)}_{j},x_2),
\label{tK}
\\
&\A_{ij}=\phi_{0,4N}(x^{(0)}_i,x^{(4N)}_j).
\label{A}
\end{align}
Hence, we find that in general class of 
measures, which is described by products of determinant,
the correlation function is represented as the determinant. 

Noticing the Jacobi-Trudi identity~\cite{St1999},
\begin{equation}
s_{\lambda/\mu}(a)=\det\left(h_{\lambda_i-\mu_j+j-i}(a)\right),
\label{JaTr}
\end{equation}
where $h_{k}(a)$ is the $k$th complete symmetric 
polynomial in variables $a=(a_1,a_2,\cdots)$,
\begin{equation}
h_{k}(a)=\sum_{i_1\le\cdots\le i_k}a_{i_1}\cdots a_{i_k},
\label{csf} 
\end{equation} 
we easily find that the weight~\eqref{pschurp} has the form 
of \eqref{pdet}, i.e., products of determinants
under the following identification:
\begin{align}
\label{i1}
&x^{(0)}=(0,-1,-2,\cdots),\\
&x^{(4N)}=(m_1,m_2,\cdots,m_n,-n,-n-1,\cdots),
\label{i2}
\\
&m_i=\mu^{(4N)}_i-i+1,
\label{mu4N}
\\
&x^{(j)}=(\lambda^{(j)}_1,\lambda^{(j)}_2-1,\lambda^{(j)}_3-2,\cdots),~~
\text{for~} j=1,\cdots, 4N-1,\\
&\phi_{r,r+1}(x^{(r)}_i,x^{(r+1)}_j)=
\begin{cases}
h_{x^{(r+1)}_i-x^{(r)}_j}(a^{(r+1)}), & \text{for~} r\text{~even},\\
h_{x^{(r)}_i-x^{(r+1)}_j}(a^{(r+1)}), & \text{for~} r\text{~odd}.
\end{cases}
\label{i4}
\end{align}
Note that in~\eqref{JaTr}, the rank of the determinant can be 
infinite (infinitely many walkers can move); hence, in this case,
$M$ in~\eqref{pdet} and~\eqref{tK} is infinity. 
Thus, under this identification~\eqref{i1}-\eqref{i4} 
with $M\rightarrow\infty$, the result~\eqref{rcor}-\eqref{A} is
applicable to case~\eqref{pschurp}.
\subsection{Double integral formula}\label{dint}
One of the purposes of this paper is to obtain
a double integral formula of the correlation kernel in 
case~\eqref{pschurp},
which is useful for the analysis of the scaling limit
as $N$ goes to infinity. 
In case of the Schur process~\eqref{schurp}, the integral formula
was first obtained in~\cite{OkRe2003} by using the fermion 
operators. In this paper, however, for considering the 
case~\eqref{pschurp}, we generalize the approach in~\cite{Jo2003} 
of calculating the inverse of the matrix  
 $\A$ in~\eqref{tK} in case~\eqref{i1}--\eqref{i4}.
In the case of the Schur process~\eqref{schurp}, i.e., 
the case $\mu=\phi$ in~\eqref{mu4N}, the matrix $\A$ is a Toeplitz 
matrix and its inverse $\A^{-1}$ can be estimated by using the 
Wiener-Hopf factorization~\cite{Jo2003}. 
However, in case~\eqref{pschurp}, where $\mu^{(4N)}$ is general, 
we cannot apply the method because $\A$ is
not a Toeplitz matrix anymore, which is the main difficulty 
of this problem. 

In this paper, we develop the method of estimating $\A^{-1}$
in the case of the weight~\eqref{pschurp} and 
give the double integral formula for the correlation kernel.
In the following theorem, we consider the situation
\begin{align}
a^{(i)}=(a^{(i)}_1,a^{(i)}_2,\cdots, a^{(i)}_p,0,0,0,\cdots),
\label{0a1}
\end{align}
where $0<a^{(i)}_j<1~~\text{for~~} i=1,\cdots,4N-1$ and $j=1,\cdots,p$.
We can easily find that the conditions above ensures the convergence 
of the partition function $Z$ in~\eqref{cor}. 

The theorem is summarized as follows. The proof is 
given in the next section.

\begin{theorem}\label{th1}
In the case~\eqref{i1}--\eqref{0a1} and $r_i=2u_i~(i=1,2)$ 
in~\eqref{kernel}, $\tilde{K}(2u_1,x_1;2u_2,x_2)$
and $\phi_{2u_1,2u_2}(x_1,x_2)$ in the correlation kernel~\eqref{kernel} 
become
\begin{align}
&~\tilde{K}(2u_1,x_1;2u_2,x_2)\notag\\
&=\frac{1}{(2\pi i)^2}\int_{C_{r_1}} 
\frac{dz_1}{z^{1+x_1}_1}
\int_{C_{r_2}}\frac{dz_2}{z^{1-x_2}_2}
\prod_{m=1}^p \frac{\prod_{i=1}^{u_1}(1-a^{(2i-1)}_m/z_1)
\prod_{j=u_2+1}^{2N}(1-a^{(2j)}_mz_2)}
{\prod_{k=u_1+1}^{2N}(1-a^{(2k)}_mz_1)\prod_{\ell=1}^{u_2}
(1-a^{(2\ell-1)}_m/z_2)}\notag\\
&~\times\left(\frac{z_1}{z_1-z_2}
+\frac{1}{s_{\mu^{(4N)}}(a)}\sum_{j=1}^{n}
\sum_{\ell'=0}^{m_j-1}h_{\ell'}(a)
z_1^{m_j-\ell'}\sum_{b=1}^{n}(-1)^{j+b}z_2^{b-1}
s_{\tilde{\mu}^{(j)}/\nu^{(b)}}(a)
\right),
\label{result1}\\
&\phi_{2u_1,2u_2}(x_1,x_2)=\frac{1}{2\pi i}\int_{C_1}\frac{dz}{z^{1+x_2-x_1}}
\prod_{k=1}^p\prod_{i=u_1}^{u_2-1}\frac{1}{1-a^{(2i+1)}_kz}
\prod_{i=u_1+1}^{u_2}\frac{1}{1-a^{(2i)}_k/z},
\label{result12}
\end{align}
where $C_{r_i}$ denotes the contour with radius $r_i$ 
satisfying $r_1> r_2$, $h_m(a)$ is the $m$th complete
symmetric polynomial~\eqref{csf} in variables
\begin{equation}
a=(a^{(1)}_1,a^{(3)}_1,a^{(5)}_1,\cdots,a^{(4N-1)}_1,
a^{(1)}_2,a^{(3)}_2,a^{(5)}_2,\cdots,a^{(4N-1)}_2,\cdots
,a^{(1)}_p,a^{(3)}_p,a^{(5)}_p,\cdots,a^{(4N-1)}_p), 
\label{21a}
\end{equation} 
and
\begin{align}
&\mu^{(4N)}=(\mu^{(4N)}_1,\mu^{(4N)}_2,\cdots,\mu^{(4N)}_n)
=(m_1,m_2+1,\cdots,m_n+n-1),\\
&\tilde{\mu}^{(j)}=(\mu^{(4N)}_1+1,\cdots,\mu^{(4N)}_{j-1}+1,
\mu^{(4N)}_{j+1},\cdots,\mu^{(4N)}_n),\\
&\nu^{(b)}=(\underbrace{1,1,\cdots,1}_{b-1},
\underbrace{0,\cdots,0}_{n-b+1}).
\label{21nu}
\end{align}
\end{theorem}

As was discussed in~\cite{Jo2003,BoRa2006}, 
the condition~\eqref{0a1} can be relaxed
such that the product 
$
\prod_{m=1}^p \frac{\prod_{i=1}^{u_1}(1-a^{(2i-1)}_m/z_1)
\prod_{j=u_2+1}^{2N}(1-a^{(2j)}_mz_2)}
{\prod_{k=u_1+1}^{2N}(1-a^{(2k)}_mz_1)\prod_{\ell=1}^{u_2}
(1-a^{(2\ell-1)}_m/z_2)}
$
converges.
Note that in case $m_i=0$(the Schur process), the second term
in~\eqref{result1} vanishes and it reduces to the integral formula
in~\cite{OkRe2003}.

\subsection{Edge scaling limit}
The benefit of the representation~\eqref{result1} is
that we can take an asymptotic 
limit of the correlation kernel. In this paper, we consider the edge scaling 
limit of the correlation kernel in the special situation 
\begin{align}
\label{set}
&x^{(4N)}=(m,-1,-2,\cdots),\\
\label{set3}
&a^{(1)}=\cdots=a^{(4N)}=(\a,0,0,\cdots).
\end{align}
In~\eqref{set} with $m=0$, 
the scaling limit of the correlation function was 
analyzed in~\cite{Jo2003} in order to study the 
height fluctuation of the PNG model.
We discuss how the scaling behavior depends on the value of
$m$.

It has been known that in the case $m=0$, 
the trace of the first particle 
$x_1^{(t)},~~t=0,\cdots,4N,$ behaves asymptotically
like
\begin{equation}
\lim_{N\rightarrow\infty}\frac{x_1^{(t)}}{N}= 
A(t):=\frac{2\a^2}{1-\a^2}+\frac{\a}{1-\a^2}
\sqrt{4t/N-(t/N)^2}.
\label{defA}
\end{equation}
This represents a semi-circle centered at $t=2N$.
We focus our attention on the fluctuation of the
walkers' positions $x_i,~i=1,2,\cdots,$
around the above limiting value at $t=2N$.  
The scaling limit is called the edge scaling 
limit. Precisely, $x_i$ and $2u_i$ in~\eqref{result1} 
and~\eqref{result12} are scaled as follows:
\begin{align}
\label{esl2}
&x_i=A(2u_i)N+DN^{1/3}\xi_i,\\
&2u_i=2N+2CN^{2/3}\t_i,
\label{esl}
\end{align}
where $A(t)$ is defined in~\eqref{defA} and
\begin{align}
&D=\frac{\a^{1/3}}{1-\a^2}(1+\a)^{4/3},~~
C=\frac{(1+\a)^{2/3}}{\a^{1/3}}.
\label{ADC}
\end{align}
The exponent $1/3$ (resp.$2/3$) in~\eqref{esl2} (resp.~\eqref{esl})
characterizes the one-dimensional KPZ universality~\cite{Jo2005L,Sp2006,
FePr2006,Ma2006,Sa2007}.
We also scale $m$ in \eqref{set} as
\begin{equation}
m=A(2N)N+BN^{2/3}\o,
\label{eslm}
\end{equation}
where
\begin{equation}
B=\frac{2\a^{2/3}}{(1-\a)(1+\a)^{1/3}}.
\label{esB}
\end{equation}

Our result of the scaling limit is summarized as follows.
The proof will be given in Section~\ref{asan}.
\begin{theorem}\label{th2}
In the situation~\eqref{set}--\eqref{esB}, the correlation kernel has the 
following scaling limit
\begin{align}
&~\lim_{N\rightarrow\infty}K(2u_1,x_1;2u_2,x_2)
\frac{DN^{1/3}}{P}\notag\\
&=
\begin{cases}
\K_2(\t_1,\xi_1;\t_2,\xi_2)
+\int_{0}^\infty d\lambda e^{\lambda(\t_1+\o)} \Ai (\xi_1-\lambda)\Ai (\xi_2), 
& \t_1+\o\le 0,\\
\K_2(\t_1,\xi_1;\t_2,\xi_2)
-\int_{0}^\infty d\lambda e^{-\lambda(\t_1+\o)} \Ai (\xi_1+\lambda)\Ai (\xi_2)\\
+\Ai(\xi_2)\exp\left(-\frac{(\t_1+\o)^3}{3}+\xi_1(\t_1+\o)\right),
&\t_1+\o> 0,
\end{cases}
\label{result2}
\end{align}
where $D$ is defined in~\eqref{ADC}, $P$ (see~\eqref{prefactor}) 
is a factor which does not contribute the determinant in~\eqref{rcor}, 
and
\begin{align}
\K_2(\t_1,\xi_1;\t_2,\xi_2)=
\begin{cases}
\int_{0}^{\infty}d\nu e^{-\nu(\t_1-\t_2)}\Ai (\xi_1+\nu)\Ai (\xi_2+\nu),
& \t_1\ge \t_2,\\
-\int_{-\infty}^{0}d\nu e^{-\nu(\t_1-\t_2)}\Ai (\xi_1+\nu)\Ai (\xi_2+\nu),
& \t_1< \t_2.
\end{cases}
\label{extAi}
\end{align}
\end{theorem}
When $m=0$ in~\eqref{set}, which corresponds to case
\eqref{schurp}($\mu^{(4N)}=\phi$), only the term
$\K_2(\t_1,\xi_1;\t_2,\xi_2)$ is left since this case 
corresponds to $\o\rightarrow -\infty$. 
The correlation kernel $\K_2(\t_1,\xi_1;\t_2,\xi_2)$ is called
the extended Airy kernel, which
appeared in the random matrix theory \cite{Fo1993,TrWi1994,Ma1994,
FoNaHo1999} and the PNG model~\cite{Jo2003,PrSp2002b}. Recently, 
the correlation kernel~\eqref{result2} has also appeared in various 
fields such as
the random matrix with external source (including the noncolliding
Brownian motion with pinned initial or final condition)~\cite{ImSa2005,
AdDeMo2007p}, the PNG model with an external source~\cite{Fo2000p,ImSa2004}, 
statistics~\cite{BaBePe2005}, and so on.

\section{Integral representation for the correlation kernel}\label{proof1}
In this section, we give the proof of Theorem~\ref{th1}.
First, we represent $\phi_{r,r+1}(x,y)$~\eqref{i4} in terms of the
vertex operators defined in~\eqref{avo}. 
The analysis of the Schur process using the vertex operators were
developed in~\cite{OkRe2003,Ok1999}. However, note that in our 
analysis, we adopt the one-particle basis whereas 
in~\cite{OkRe2003,Ok1999}, infinitely-many particle state is 
chosen as a basis. Basic properties of the vertex operators 
and symmetric functions are summarized in Appendix A. 

First, we prove~\eqref{result12}. 
From~\eqref{JaTr} and~\eqref{aph}, we have
\begin{equation}
\phi_{r,r+1}(x,y)=
\begin{cases}
\langle y|H_{+}(s^{(r+1)})|x\rangle , & \text{for~} r\text{~even},\\
\langle y|H_{-}(s^{(r+1)})|x\rangle, & \text{for~} r\text{~odd}.
\end{cases}
\label{phxy}
\end{equation}
Here, the parameter $s^{(j)}=(s^{(j)}_1,s^{(j)}_2,\cdots)$ 
is connected to $a^{(j)}$ in~\eqref{pschurp} as
\begin{equation}
s^{(j)}_k=\frac{1}{k}\sum_{i=1}^p a_i^{(j)k}.
\label{palsa}
\end{equation}
By using~\eqref{phxy} and~\eqref{ap4}, we can express 
$\phi_{2u_1,2u_2}$~\eqref{dphi} as 
\begin{equation}
\phi_{2u_1,2u_2}(x_1,x_2)=
\langle x_2|
\prod_{i=u_1}^{u_2-1}H_+(s^{(2i+1)})
\prod_{j=u_1+1}^{u_2}H_-(s^{(2j)})
|x_1\rangle .
\label{vphi}
\end{equation}
By applying~\eqref{ap1} and \eqref{ap2} to this equation,
we easily get \eqref{result12}.

Next, we consider the integral form of $\tilde{K}$.
Eq.~\eqref{tK} is written as
\begin{align}
&~\tilde{K}(2u_1,x_1;2u_2,x_2)\notag\\
&=\frac{1}{(2\pi i)^2}\int_{C_{r_1}}\frac{dz_1}{z_1^{1+x_1}}
\int_{C_{r_2}}\frac{dz_2}{z_2^{1-x_2}}
\frac{z_1}{z_2}\sum_{i,j=1}^{\infty}z_1^{x_i^{(4N)}-1}\A^{-1}_{ij}
z_2^{x^{(0)}_j+1}
\left(\sum_{a=-\infty}^{\infty}\phi_{2u_1,4N}(a,x_i^{(4N)})
z_1^{a-x_i^{(4N)}}\right)
\notag\\
&~~~~\times
\left(\sum_{b=-\infty}^{\infty}\phi_{0,2u_2}(x^{(0)}_j,b)
z_2^{x^{(0)}_j-b}\right)\notag\\
&=\frac{1}{(2\pi i)^2}\int_{C_{r_1}}\frac{dz_1}{z_1^{1+x_1}}
\int_{C_{r_2}}\frac{dz_2}{z_2^{1-x_2}}
\frac{z_1}{z_2}\sum_{i,j=1}^{\infty}z_1^{x_i^{(4N)}-1}\A^{-1}_{ij}z_2^{j}
\notag\\
&~~~~\times\prod_{i=u_1}^{2N-1}\g(z_1^{-1},s^{(2i+1)})
\prod_{j=u_1+1}^{2N}\g(z_1,s^{(2j)})
\prod_{k=0}^{u_2-1}\g(z_2^{-1},s^{(2k+1)})
\prod_{\ell=1}^{u_2}\g(z_2,s^{(2\ell)}).
\label{tKi}
\end{align}
Here, $\g(z,s)$ is defined in~\eqref{agen} and $C_{r_i}~(i=1,2)$ 
denotes the contour centered at the origin anticlockwise 
with radius $r_i$ satisfying $r_2<r_1$.
In the second equality, we use~\eqref{vphi} 
and~\eqref{ap1}--\eqref{ap3}.

Thus, what we have to do is to obtain an explicit form of 
the inverse of the (semi-infinite) matrix $\A$~\eqref{A} 
in case~\eqref{i1}--~\eqref{i4} and calculate
$\sum_{i,j=1}^{\infty}z_1^{x_i^{(4N)}-1}\A^{-1}_{ij}z_2^{j}$.
From~\eqref{vphi}, the matrix $\A$ in our case~\eqref{i1}--\eqref{i4}
can be expressed as  
\begin{equation}
\A_{ij}=\langle x_j|\L_+(s)\L_-(s')|1-i\rangle
\label{vA},
\end{equation}
where
\begin{align}
\label{3m}
&x_j(=x_j^{(4N)})=
\begin{cases}
m_j, & \text{for~} 1\le j\le n,\\
1-j, & \text{for~} n+1\le j, 
\end{cases}\\
\label{3L}
&\L_{+}(s)=\prod_{j=1}^{2N}H_{+}(s^{(2j-1)}),~~
\L_{-}(s')=\prod_{j=1}^{2N}H_{-}(s^{(2j)}),
\\
&s=(s^{(1)},s^{(3)},\cdots,s^{(4N-1)}),~~
s'=(s^{(2)},s^{(4)},\cdots,s^{(4N)}).
\end{align}

In order to get $\A^{-1}$, 
we introduce the following matrix:
\begin{equation}
\A'_{ij}=\langle 1-j|\L^{-1}_-(s')\P'\L^{-1}_+(s)|x_i\rangle,
\end{equation}
where $\P'$ is a deformed projection operator
\begin{equation}
\P'=\sum_{j=1}^\infty|1-j\rangle\langle x_j|.
\end{equation}
Note that when $n=0$ in~\eqref{3m}, $\P'$ becomes an ordinary
projection operator $\P$ defined as
\begin{equation}
\P=\sum_{j=1}^\infty|1-j\rangle\langle 1-j|.
\end{equation} 
By the use of $\P'$ and $\P$, we can express the product
$\A'\A$ as
\begin{align}
(\A'\A)_{ik}
=\langle x_k|\L_+(s)\L_-(s')\P\L_-^{-1}(s')\P'\L_+^{-1}(s)|x_i\rangle.
\label{AA'}
\end{align}
Let us focus on the term
\begin{equation}
\P\L_-^{-1}(s')\P'=\sum_{p=1}^{\infty}\sum_{p'=1}^{\infty}
|1-p\rangle\langle 1-p|\L_-^{-1}(s')|1-p'\rangle\langle x_{p'}|.
\label{plpd}
\end{equation} 
Noticing $1-p'<0$ and $<j|\L_-^{-1}(s')|i>=0$ for $j>i$, which
immediately follows from definition~\eqref{3L}, we 
have
\begin{equation}
\P\L_-^{-1}(s')\P'=\L_-^{-1}(s')\P'.
\label{PI}
\end{equation}
Namely, we can change $\P$ into 
the identity operator in~\eqref{plpd}. Thus, we obtain
\begin{align}
(\A'\A)_{ik}=\langle x_k|\L_+(s)\P'\L_+^{-1}(s)|x_i\rangle.
\label{3ada}
\end{align}
Note that when $n=0$, one can easily find that $\P'(=\P)$ can also
be changed into the identity operator in~\eqref{3ada} from the similar 
discussion
about~\eqref{PI}. Thus, in the case $n=0$, $\A'\A=1$ or, equivalently,
$\A'$ is nothing but $\A^{-1}$.

In the case of general $n$, we can represent the matrix $\A'\A$ 
in terms of $c_{j-k}(s)$ and $d_{j-k}(s)$ defined 
in~\eqref{ac} and~\eqref{ad} respectively, i.e.,
\begin{align}
c_{j-k}(s)=\langle j|\L_+(s)|k\rangle
=\langle k|\L_-(s)|j\rangle,~~
d_{j-k}(s)=\langle j|\L_+^{-1}(s)|k\rangle
=\langle k|\L_-^{-1}(s)|j\rangle.
\end{align}
By noting $c_{j-k}=d_{j-k}=0$ for $j-k<0$, we get
\begin{equation}
\A'\A=
\begin{pmatrix}
a_{11}&a_{12}&\cdots&a_{1n}&&&\\
a_{21}&a_{22}&\cdots&a_{2n}&&&\\
\vdots&\vdots&&\vdots&&&\\
a_{n1}&a_{n2}&\cdots&a_{nn}&&&\\
a_{n+11}&a_{n+12}&\cdots&a_{n+1n}&1&&\\
\vdots&\vdots&&\vdots&&1&\\
\vdots&\vdots&&\vdots&&&\ddots
\end{pmatrix},
\end{equation}
where the region after the $n+1$th column is equivalent to the 
unit matrix and
\begin{equation}
a_{ik}=
\begin{cases}
\sum_{j=1}^i d_{m_j-m_i}(s)c_{m_k+j-1}(s), & 1\le i\le n,~1\le k\le n,\\
\sum_{j=n+1}^ic_{m_k+j-1}(s)d_{i-j}(s)+\sum_{j=1}^nd_{m_j+i-1}(s)
c_{m_k+j-1}(s),
&n+1\le i,~1\le k\le n.
\end{cases}
\label{aij}
\end{equation}
Here, $m_i$ is defined in~\eqref{mu4N}.
Note that the matrix $\A'\A$ is almost the unit matrix. 
We easily find that 
the inverse of the matrix $\A'\A$ has the following form:
\begin{align}
\B&:=(\A'\A)^{-1}=\frac{1}{b_0}
\begin{pmatrix}
b_{11}&b_{12}&\cdots&b_{1n}&&&\\
b_{21}&b_{22}&\cdots&b_{2n}&&&\\
\vdots&\vdots&&\vdots&&&\\
b_{n1}&b_{n2}&\cdots&b_{nn}&&&\\
b_{n+11}&b_{n+12}&\cdots&b_{n+1n}&b_0&&\\
\vdots&\vdots&&\vdots&&b_0&\\
\vdots&\vdots&&\vdots&&&\ddots
\end{pmatrix}.
\end{align}
By introducing 
\begin{align}
&\A'':=(a_{ij})_{i,j=1}^n,~~\tilde{\A''}_{(ji)}: 
~ji~\text{cofactor of~~} \A'',\\ 
&\C=(c_{m_j+i-1}(s))_{i,j=1}^n,
\label{matc}
\end{align}
we find that the elements $b_0$ and $b_{i,j}$ can be expressed 
as
\begin{align}
&b_0=\det \A''=\det \C, \\
&b_{ij}=
\begin{cases}
\tilde{\A''}_{(ji)},&1\le i\le n,~1\le j\le n\\
-\det
\begin{pmatrix}
a_{11}&a_{12}&\hdots&a_{1n}\\
\vdots&\vdots&&\vdots\\
a_{j-11}&a_{j-12}&\hdots&a_{j-1n}\\
a_{i1}&a_{i2}&\hdots&a_{in}\\
a_{j+11}&a_{j+12}&\hdots&a_{j+1n}\\
\vdots&\vdots&&\vdots\\
a_{n1}&a_{n2}&\hdots&a_{nn}
\end{pmatrix}
=-\sum_{k=1}^na_{ik}b_{kj},
&n+1\le i,~1\le j\le n
\end{cases}.
\label{bij}
\end{align}

Thus the inverse of the matrix $\A$~\eqref{A} is 
represented as
\begin{align}
\A^{-1}
&=\B\A'\notag\\
&=
\begin{pmatrix}
\frac{1}{b_0}\sum_{k=1}^nb_{1k}\A'_{k1}&
\frac{1}{b_0}\sum_{k=1}^nb_{1k}\A'_{k2}&\hdots\\
\vdots&\vdots&\\
\frac{1}{b_0}\sum_{k=1}^nb_{nk}\A'_{k1}&
\frac{1}{b_0}\sum_{k=1}^nb_{nk}\A'_{k2}&\hdots\\
\frac{1}{b_0}\sum_{k=1}^nb_{n+1k}\A'_{k1}+\A'_{n+1,1}&
\frac{1}{b_0}\sum_{k=1}^nb_{n+1k}\A'_{k2}+\A'_{n+1,2}&\hdots\\
\frac{1}{b_0}\sum_{k=1}^nb_{n+2k}\A'_{k1}+\A'_{n+2,1}&
\frac{1}{b_0}\sum_{k=1}^nb_{n+2k}\A'_{k2}+\A'_{n+2,2}&\hdots\\
\vdots&\vdots&\\
\end{pmatrix}.
\end{align}
By using this expression for $\A^{-1}$, we can represent 
$\sum_{i,j=1}^{\infty}z_1^{x_i^{(4N)}-1}\A^{-1}_{ij}z_2^{j}$
in~\eqref{tKi} as
\begin{align}
\sum_{i,j=1}^\infty z_1^{x_i-1}(\A^{-1})_{ij}z_2^j
&=
\sum_{i=n+1}^{\infty}\sum_{j=1}^\infty z_1^{-i}\A'_{ij}z_2^{j}
+\frac{1}{b_0}\sum_{k=1}^n\sum_{i=n+1}^\infty z_1^{-i}b_{ik}
\sum_{j=1}^\infty \A'_{kj}z_2^{j}\notag\\
&~~+\frac{1}{b_0}\sum_{k=1}^n\sum_{i=1}^n z_1^{m_i-1}b_{ik}
\sum_{j=1}^\infty \A'_{kj}z_2^{j}.
\label{zwa-1}
\end{align}
After quite lengthy calculations of this equation, 
we obtain the following expression
\begin{align}
\sum_{i,j=1}^\infty z_1^{x_i-1}(\A^{-1})_{ij}z_2^j
&=
\left(\sum_{k=1}^{\infty}\left(\frac{z_2}{z_1}\right)^{k}
+
\frac{1}{\det\C}\sum_{j=1}^n\sum_{\ell'=0}^{m_j-1}c_{\ell'}(s)
z_1^{m_j-1-\ell'}\sum_{b=1}^{n}z_2^{b}\tilde{\C}_{(bj)}\right)
\notag\\
&~~\times\prod_{j=1}^{2N}\g^{-1}(z_1^{-1}, s^{(2j-1)})
\g^{-1}(z_2, s^{(2j)}),
\label{apB}
\end{align}
where $\tilde{\C}_{(bj)}$ is the $bj$ cofactor of matrix $\C$.
The proof of this equation is given in Appendix B.
By noticing~\eqref{aph} and~\eqref{ac}, 
we can rewrite $c_{\ell'}(s)$, $\det\C$, and 
$\tilde{\C}_{(bj)}$ as
\begin{align}
\label{3apb1}
&c_{\ell'}(s)=h_{\ell'}(a),\\
&\det\C=s_{\mu^{(4N)}}(a),\\
&\tilde{\C}_{(bj)}=(-1)^{j+b}s_{\tilde{\mu}^{(j)}/\nu^{(b)}}(a),
\label{3apb3}
\end{align} 
where $a,~\mu^{(4N)},~\tilde{\mu}^{(j)}$, and $\nu^{(b)}$ 
are defined in~\eqref{21a}-\eqref{21nu}, respectively.

By substituting~\eqref{apB}-\eqref{3apb3} into~\eqref{tKi} and 
noting~\eqref{palsa} and~\eqref{agen}, we get the desired 
expression~\eqref{result1}.

\section{Asymptotic analysis}\label{asan}
In this section, we give the proof of Theorem~\ref{th2}.
We investigate the asymptotics of the correlation kernel
by the use of the saddle point analysis. 

In case~\eqref{set} and \eqref{set3}, the correlation kernel defined
in~\eqref{result1} and~\eqref{result12} reduces to 
\begin{align}
&~K(2u_1,x_1;2u_2,x_2)=\tilde{K}(2u_1,x_1;2u_2,x_2)
-\phi_{2u_1,2u_2}(x_1,x_2)
\notag\\
&=\frac{1}{(2\pi i)^2}\int_{C_{r_1}} \frac{dz_1}{z^{1+x_1}_1}
\int_{C_{r_2}}\frac{dz_2}{z^{1-x_2}_2}
\frac{z_1}{z_1-z_2}
\frac{(1-\a/z_1)^{u_1}(1-\a z_2)^{2N-u_2}}
{(1-\a z_1)^{2N-u_1}(1-\a/z_2)^{u_2}}\notag\\
&+\frac{1}{2\pi i}\int_{C_1}\frac{dz}{z^{1+x_2-x_1}}
(1-\a z)^{u_1-u_2}(1-\a/z)^{u_1-u_2}\notag\\
&+
\sum_{j=1}^{m}
\frac{h_{m-j}(\a^{2N})}{h_m(\a^{2N})}
\frac{1}{2\pi  i}\int_{C_{1}}\frac{dz_1}{z_1^{1+x_1-j}}
\frac{(1-\a/z_1)^{u_1}}{(1-\a z_1)^{2N-u_1}}\times
\frac{1}{2\pi i}\int_{C_{1}}\frac{dz_2}{z_2^{1-x_2}}
\frac{(1-\a z_2)^{2N-u_2}}{(1-\a/z_2)^{u_2}},
\end{align}
where $C_{r}$ denotes the contour enclosing the origin 
anticlockwise with radius
$r$ and $h_j(\a^{2N})$ is the $j$th complete symmetric 
polynomial~\eqref{csf} in variables
\begin{equation}
\a^{2N}=\underbrace{(\a,\a,\cdots,\a)}_{2N}.
\end{equation}

In~\cite{Jo2003}, it was shown that the first two terms of the correlation 
kernel
above converge to the extended Airy kernel in the edge scaling 
limits~\eqref{esl2} and \eqref{esl},
\begin{align}
&~~\lim_{N\rightarrow\infty}\frac{DN^{1/3}}{(2\pi i)^2}
\int_{C_{r_1}} \frac{dz_1}{z^{1+x_1}_1}
\int_{C_{r_2}}\frac{dz_2}{z^{1-x_2}_2}
\frac{z_1}{z_1-z_2}
\frac{(1-\a/z_1)^{u_1}(1-\a z_2)^{2N-u_2}}
{(1-\a z_1)^{2N-u_1}(1-\a/z_2)^{u_2}}\notag\\
&~~+\frac{DN^{1/3}}{2\pi i}\int_{C_1}\frac{dz}{z^{1+x_2-x_1}}
(1-\a z)^{u_1-u_2}(1-\a/z)^{u_1-u_2}
=P\times
\K_2(\t_1,\xi_1;\t_2,\xi_2),
\label{exp1}
\end{align}
where $\K_2(\t_1,\xi_1;\t_2,\xi_2)$ is defined in~\eqref{extAi} and
$P$ is a prefactor which does not contribute a determinant,
\begin{equation}
P=(1-\a)^{\frac{2(1+\a)^{2/3}N^{2/3}(\t_1-\t_2)}
{\a^{1/3}}}e^{(\t_1^3-\t_2^3)/3+\xi_2\t_2-\xi_1\t_1}.
\label{prefactor}
\end{equation}
Thus, we only consider the asymptotics of the last term,
\begin{equation}
\sum_{j=1}^{m}
\frac{h_{m-j}(\a^{2N})}{h_{m}(\a^{2N})}\psi_1(x_1-j)\psi_2(x_2),
\label{sct}
\end{equation}
where
\begin{align}
\label{psi1}
&\psi_1(x_1-j)=\frac{1}{2\pi i}\int_{C_{1}}\frac{dz_1}{z_1^{1+x_1-j}}
\frac{(1-\a/z_1)^{u_1}}{(1-\a z_1)^{2N-u_1}} ,\\
&\psi_2(x_2)=\frac{1}{2\pi i}\int_{C_{1}}\frac{dz_2}{z_2^{1-x_2}}
\frac{(1-\a z_2)^{2N-u_2}}{(1-\a/z_2)^{u_2}},
\label{psi2}
\end{align}
under the scaling~\eqref{esl2}--\eqref{esB}. We scale $j$ as
\begin{equation}
j=DN^{1/3}\lambda.
\label{jsc}
\end{equation}
Here, $D$ is defined in~\eqref{ADC}.
This scaling makes a dominant contribution to the asymptotic limit
of~\eqref{sct}.

First, we consider the function $\psi_1(x_1-j)$ .
By substituting~\eqref{esl} into~\eqref{psi1},
we have
\begin{align}
\psi_1(x_1-j)=
\frac{1}{2\pi i}\int_{C_1}\frac{dz_1}{z_1^{1+x_1-j-\mu N}}
\exp(Nf_{\b,\mu}(z_1)),
\label{4psi}
\end{align}
where
\begin{align}
&f_{\b,\mu}(z)=(1+\b)\log(z-\a)-(1-\b)\log (1-\a z)-(\mu+1+\b)
\log z,\notag\\
&\b=CN^{-1/3}\t_1, 
\label{fmyu}
\end{align}
and we fix $\mu$ as
\begin{equation}
\mu=\frac{2\a}{1-\a^2}(\a+\sqrt{1-\b^2})=A(2u_1),
\label{mu}
\end{equation}
where $A(2u_1)$ is defined in~\eqref{defA}.
In~\eqref{fmyu}, $C$ is defined in~\eqref{ADC}.
We find that due to~\eqref{mu}, two saddle points of the function
$f_{\b,\mu}(z)$ merge to the double saddle point $z_c(\b)$,
\begin{equation}
z_c(\b)=\frac{\sqrt{1+\b}+\a\sqrt{1-\b}}{\sqrt{1-\b}+\a\sqrt{1+\b}}.
\end{equation}
We also have the relations
\begin{align}
&\frac{f_{\b,\mu}(z_c(\b))}{dz}
=\frac{d^2f_{\b,\mu}(z_c(\b))}{dz^2}=0,~
\frac{d^{3}f_{\b,\mu}(z_c(\b))}{dz^3}
%=-\frac{2\a(1+\a)}{(1-\a)^3}
=\frac{2D^3}{z^3_c(\b)}.
\end{align}
Since the main contribution to the integral in~\eqref{4psi} 
is given around $z\sim z_c(\b)$, $z_1$ may be transformed to
\begin{align}
z_1&=z_c(\b)\left(1-\frac{iw_1}{DN^{1/3}}\right)\sim
1+\frac{\t_1-iw_1}{DN^{1/3}}.
\label{z1w1}
\end{align}
Then, we obtain the relations
\begin{align}
\exp(Nf_{\b,\mu}(z_1))&\sim\exp\left(N\left(f_{\b,\mu}(z_c(\b))
+\frac{f_{\b,\mu}'''(z_c(\b))}{6}(z-z_c(\b))^3
\right)\right)\notag\\
&=\exp(Nf_{\b,\mu}(z_c(\b)))
\exp\left(\frac{i}{3}w_1^3\right),\\
\frac{1}{z_1^{x_1-j+1-\mu N}}&\sim \exp((\xi_1-\lambda)(iw_1-\t_1)).
\label{prel2}
\end{align}
In~\eqref{prel2}, we used~\eqref{esl2}.
By using ~\eqref{z1w1}--~\eqref{prel2}, and
\begin{equation}
\exp\left(Nf_{\b,\mu}(z_c(\b))\right)
\sim(1-\a)^{\frac{2(1+\a)^{2/3}\t_1N^{2/3}}{\a^{1/3}}}
\exp(\t_1^3/3),
\end{equation}
we get
\begin{align}
DN^{1/3}\psi_1(x_1-j)&\sim
(1-\a)^{\frac{2(1+\a)^{2/3}N^{2/3}\t_1}
{\a^{1/3}}}\exp\left(\t_1^3/3
-\xi_1\t_1+\t_1\lambda\right)\times\Ai(\xi_1-\lambda).
\label{apsi1}
\end{align}
Similarly $\psi_2(x_2)$ becomes
\begin{align}
DN^{1/3}\psi_2(x_2)&\sim
(1-\a)^{\frac{-2(1+\a)^{2/3}N^{2/3}\t_2}
{\a^{1/3}}}\exp\left(-\t_2^3/3
+\xi_2\t_2\right)\times\Ai(\xi_2).
\label{apsi2}
\end{align}
Here, in the equations above, we used the integral representation of 
the Airy function
\begin{equation}
\Ai(x)=\int_{-\infty}^{\infty}d\lambda e^{ix\lambda+\frac{i}{3}\lambda^3}.
\end{equation}

Next, we also apply the saddle point analysis to  
$h_{m-j}(\a^{2N})/h_m(\a^{2N})$ in~\eqref{sct} under 
the scaling~\eqref{eslm} and~\eqref{jsc} as 
$N\rightarrow\infty$ .
From~\eqref{agen},~\eqref{acp2}, and~\eqref{eslm}, $h_{m-j}(\a^{2N})$ 
is expressed as
\begin{align}
h_{m-j}(\a^{2N})
=\frac{1}{2\pi i}\int_{C_1} e^{-Ng(z)}\frac{dz}{z^{-j+1}},
\label{heg}
\end{align}
where the contour $C_1$ represents the unit circle surrounding the origin 
anticlockwise, and
\begin{align}
&g(z)=2\log (1-\a z)+\left(\frac{2\a}{1-\a}+\delta\right)\log z,~~
\delta=\frac{B\o}{N^{1/3}}.
\end{align} 
The saddle point $w_{c}$ of 
$g(z)$ is
\begin{equation}
w_c=\frac{2\a+(1-\a)\delta}
{2\a+\a(1-\a)\delta}.
\label{zgc}
\end{equation}
When we deform the path of $z$ to
\begin{equation}
z=w_c\left(1+\frac{1-\a}{\a^{1/2}N^{1/2}}iw\right)\sim
1+\frac{\o}{DN^{1/3}}+\frac{1-\a}{\a^{1/2}N^{1/2}}iw,
\end{equation}
we find
\begin{align}
&\exp\left(-Ng(z)\right)\sim 
\exp(-Ng(w_c)-w^2),~~\frac{1}{z^{-j+1}}\sim e^{\o\lambda}.
\end{align}
Therefore, the asymptotic form of $h_{m-j}(\a^{2N})$ becomes
\begin{align}
h_{m-j}(\a^{2N}) \sim \frac{N^{-1/2}}{2\sqrt{\pi}}\frac{1-\a}{\a}
(1-\a)^{-2N}\exp\left(-\frac{\a^{1/3}\o^2N^{\frac13}}{(1+\a)^{2/3}}
+\o^3/3+\o\lambda\right).
\label{asm}
\end{align}
Thus, we get
\begin{equation}
\frac{h_{m-j}(\a^{2N})}{h_m(\a^{2N})}\sim \exp(\lambda\o).
\label{ratio}
\end{equation}
From~\eqref{apsi1},~\eqref{apsi2}, and~\eqref{ratio}, we obtain 
\begin{align}
DN^{1/3}\sum_{j=1}^{m}
\frac{h_{m-j}(\a^{2N})}{h_{m}(\a^{2N})}\psi_1(x_1-j)\psi_2(x_2)
\sim
P\times\int_{0}^{\infty}d\lambda e^{\lambda(\t_1+\o)} \Ai(\xi_1-\lambda)\Ai (\xi_2).
\label{exp2}
\end{align}
Here the prefactor $P$ is defined in~\eqref{prefactor}.
From~\eqref{exp1} and~\eqref{exp2} and noting that $P$ does not
contribute to determinant calculation, we finally obtain the 
desired expression. However, it is obvious that this expression 
is valid only for the case $\t_i+\o<0 (i=1,2)$ because 
in the case $\t_i+\o>0$, the integration in~\eqref{exp2} is 
divergent. 

In order to get the asymptotic form in the latter 
case, we need another expression for~\eqref{sct}. 
By using~\eqref{acp2}, we rewrite the term 
$\sum_{j=1}^m h_{m-j}(\a^{2N})z^{-j}$ in~\eqref{sct},
\begin{equation}
\sum_{j=1}^mh_{m-j}(\a^{2N})z^j=\frac{z^{m}}{(1-\a/z)^{2N}}
-\sum_{j=-\infty}^{0}h_{m-j}(\a^{2N})z^j.
\end{equation}
Thus, we obtain
\begin{equation}
\sum_{j=1}^{m}
\frac{h_{m-j}(\a^{2N})}{h_{m}(\a^{2N})}\psi_1(x_1-j)\psi_2(x_2)
=\frac{1}{h_m(\a^{2N})}\varphi(x_1)\psi_2(x_2)-\sum_{j=-\infty}^{0}
\frac{h_{m-j}(\a^{2N})}{h_m(\a^{2N})}
\psi_1(x_1-j)\psi_2(x_2),
\end{equation}
where
\begin{equation}
\varphi(x_1)=\frac{1}{2\pi i}\int_{C_{1}}
\frac{dz_1}{z_1^{1+x_1-m}}(1-\a/z_1)^{u_1-2N}(1-\a z_1)^{u_1-2N},
\end{equation}
and $\psi_1(x)$(resp. $\psi_2(x)$) is defined in~\eqref{psi1}
(resp.~\eqref{psi2}).

We can easily get the asymptotic forms of $\varphi(x)$ by using 
the saddle point analysis in a manner similar to the derivation 
of~\eqref{asm}. The result is
\begin{align}
&~\varphi(x_1)\notag\\
&\sim
\frac{N^{-1/2}}{2\pi^{1/2}}\frac{1-\a}{\a}(1-2N)^{-2N}
(1-\a)^{\frac{2(1+\a)^{2/3}\t_1N^{2/3}}{\a^{1/3}}}
\exp\left(-\frac{\a^{1/3}\o^2N^{1/3}}{(1+\a)^{2/3}}
-\t_1^2\o-\t_1\o^2+\xi_1\o\right).
\label{varphi}   
\end{align}  
From~\eqref{apsi2},~\eqref{asm}, and~\eqref{varphi}, we have
\begin{equation}
DN^{1/3}\frac{\varphi(x_1)}{h_m(\a^{2N})}\psi_2(x_2)
\sim
P\times e^{-(\o+\t_1)^3/3+\xi_1(\o+\t_1)}
\Ai(\xi_2).
\end{equation}  
Analogous to the derivation of~\eqref{exp2}, we also have
\begin{align}
-DN^{1/3}\sum_{j=-\infty}^{0}
\frac{h_{m-j}(\a^{2N})}{h_{m}(\a^{2N})}\psi_1(x_1-j)\psi_2(x_2)
\sim
-P\times\int_{0}^{\infty}d\lambda e^{-\lambda(\t_1+\o)} \Ai(\xi_1+\lambda)\Ai (\xi_2).
\label{mexp2}
\end{align}
From these two relations, we get the desired expression also
in the case $\t_i+\o>0$.

\section{Conclusion}\label{con}
In this paper, we have studied the correlation function~\eqref{cor}
of the process~\eqref{pschurp}. This process is the generalization
of the Schur process~\eqref{schurp}~\cite{OkRe2003} in the sense 
that in the picture 
of nonintersecting random walk, the walkers end at fixed sites,
as depicted in Fig.~2. At first, we have obtained the integral representation 
of the correlation kernel (Theorem~\ref{th1}). 
The result is the generalization of the one in the Schur process obtained
in~\cite{OkRe2003,Jo2003,BoRa2006}.
The derivation is given in 
Sec.~\ref{proof1}. The key technique is the calculation of the inverse of 
the matrix $\A$~\eqref{A} 
with semi-infinite rank. This technique can be 
regarded as the generalization of the one discussed in~\cite{Jo2003}.
Next, by using the integral representation, we have obtained the edge 
scaling limit of the correlation kernel in the special case~\eqref{set} and 
\eqref{set3}. The result is summarized as Theorem~\ref{th2} and the 
proof is given in Sec.~\ref{asan}. We have found that the limiting 
correlation kernel is equivalent to the one obtained in other determinantal 
processes such as the random matrix with external source, the PNG
model, and statistics.

We list the future problems as follows.
\begin{enumerate}
\item In this paper, we have discussed the asymptotics of the correlation kernel 
only in the simple case~\eqref{set} and~\eqref{set3}, although the 
process~\eqref{pschurp} has many parameters $\mu^{(4N)}$ and
 $\{a^{(i)}\}_{i=1,\cdots,4N}$. It would be interesting 
to investigate the limiting behavior of the correlation kernel in 
a more general situation and how it depends on these parameters.   

\item The correlation kernel may be closely related to the orthogonal 
polynomials. In the Schur process, the correlation kernel (the first term 
in~\eqref{result1}) can be represented in terms of the Meixner polynomial
in the single time case~\cite{Jo2000}.
On the other hand, it has been recently revealed that the correlation kernel 
corresponding to the random Hermitian matrix with external source 
can be expressed in terms of the multiple orthogonal 
polynomials~\cite{BlKu2004}. 
In the discretized process~\eqref{pschurp}, is the correlation 
kernel related to any discrete analog of the multiple orthogonal polynomial?

\item In Theorem 2.1, the second term in~\eqref{result1}, 
which is due to the partition $\mu^{(4N)}$ 
in~\eqref{pschurp}, can be expressed by using the Schur function.
This fact raises our hope that there exists some deep connection 
between the process~\eqref{pschurp} and the theory of integrable 
system and representation theory. In particular, it is interesting
to view the problem in perspective of the Kadomtsev-Petviashvili 
and Toda Lattice hierarchies~\cite{AdvM2001,HaOr2007p}.
Recently, the relationship between the stochastic processes such as 
the random turn walk and soliton theory has been discussed 
in~\cite{HaOr2007p}. This approach may be useful for studying 
this topic.
\end{enumerate}

\appendix
\section{Vertex operators and symmetric functions}
Let $c^{\dagger}_j$ and $c_i~(i,j\in\Z=(\cdots,-1,0,1,\cdots))$ 
be the creation and annihilation operators
that satisfy the fermion anticommutation relation
\begin{align}
&\{c_i,c_j^{\dagger}\}:=c_ic_j^{\dagger}+c_j^{\dagger}c_i=
\delta_{ij},\\
&\{c_i,c_j\}=\{c_i^{\dagger},c_j^{\dagger}\}=0.
\end{align}
One particle state $|i\rangle$ is defined as
\begin{equation}
|i\rangle=c_i^{\dagger}|\O\rangle,
\end{equation}
where $|\O\rangle$ is the vacuum state without particle.

The mode operators $c(z),~c^{\dagger}(z)$ and the vertex operators 
$H_{\pm}(s)$, where $s=(s_1,s_2,\cdots)$, are defined as follows:
\begin{align}
&c^{\da}(z)=\sum_{i=-\infty}^{\infty}z^ic^{\da}_i,~~
c(z)=\sum_{i=-\infty}^{\infty}z^{-i}c_i,\label{cw}\\
&H_{\pm}(s)=\exp\left(\sum_{n=1}^{\infty}s_n\b_{\pm n}\right),
\text{~where~}
\b_{\pm n}=\sum_{k=-\infty}^{\infty}c^{\da}_{k\pm n}c_k,~(n=1,~2,\cdots).
\label{avo}
\end{align}
We summarize the basic properties of these operators, which are 
used for our discussion, 
\begin{align}
&H_{\pm}(s)|\O\rangle=|\O\rangle,
\label{ap1}
\\
&H_{\pm}(s)c^{\da}(z)=\g(z^{\mp 1},s)c^{\da}(z)H_{\pm}(s),
\label{ap2}
\\
&H_{\pm}(s)c(z)=\g(z^{\mp 1},s)^{-1}c(z)H_{\pm}(s),~\text{where}
~\g(z,s)=\exp\left(\sum_{n=1}^\infty s_nz^n\right),
\label{ap3}\\
&H_{+}(s)H_{-}(s')\stackrel{\langle j|\cdot|i\rangle}{=}
H_-(s')H_+(s),
\label{ap4}
\end{align}
In~\eqref{ap4}, the symbol $\stackrel{\langle j|\cdot|i\rangle}{=}$ 
means an equality on one particle space.
Note that~\eqref{ap4} is different from the ordinary one.
As discussed in~\cite{OkRe2003} when 
the product $H_{+}(s)H_{-}(s')$ appears
in another vacuum state where infinitely many particles are occupied
up to the origin, the following relation holds: 
\begin{equation}
H_{+}(s)H_{-}(s')=e^{\sum_{n=1}^{\infty}s_ns'_n}H_-(s')H_+(s).
\end{equation}

When we set the variable $s$ of the vertex operators,
\begin{equation}
s_j=\frac{1}{j}\sum_{i=1}^n a_i^{j},
\label{apar}
\end{equation}
we find that $\g(z,s)$ in~\eqref{ap3} is given as the generating
function of the complete symmetric function $h_j(a)$,
\begin{equation}
\g(z,s)=\prod_{i=1}^n\frac{1}{1-a_iz}=\sum_{j}h_j(a)z^{j}.
\label{agen}
\end{equation}
By using this and the properties~\eqref{ap1}--\eqref{ap3},
we can describe $h_j(a)$ as
\begin{equation}
h_{j-i}(a)=\langle j|H_{+}(s)|i\rangle =\langle i|H_{-}(s)|j\rangle,
\label{aph}
\end{equation}
under the parameterization~\eqref{apar}.

We also often use the following functions:
\begin{align}
&c_{j-k}(v)=\langle j|\L_+(v)|k\rangle
=\langle k|\L_-(v)|j\rangle,
\label{ac}\\
&d_{j-k}(v)=\langle j|\L_+^{-1}(v)|k\rangle
=\langle k|\L_-^{-1}(v)|j\rangle,
\label{ad}
\end{align}
where
\begin{align}
&\L_{\pm}(v)=\prod_{j=1}^{2N}H_{\pm}(v^{(j)})
%=\exp(\sum_{n\ge 1}v_n\beta_{\pm n})
,\\
&v=(v^{(1)},v^{(2)},\cdots,v^{(2N)}).
\end{align}
The properties that we often use are summarized as follows:
\begin{align}
&\sum_{i=0}^k c_{i}(v)d_{k-i}(v)=
\begin{cases}
1,&\text{for~~} k=0,\\
0,&\text{for~~} 1\le k, 
\end{cases}
\label{acp1}
\\
&\sum_{i=0}^{\infty}c_{i}(v)z^{i}=\prod_{i=1}^{2N}\g(z,v^{(i)}),~~
\sum_{i=0}^{\infty}d_{i}(v)z^{i}=\prod_{i=1}^{2N}\g^{-1}(z,v^{(i)}).
\label{acp2}
\end{align}

\section{Proof of~\eqref{apB}}
In this appendix, we derive~\eqref{apB}
by deforming the right hand side of~\eqref{zwa-1}.
By noticing
\begin{equation}
\A'_{ij}=\sum_{k=1}^{\infty}d_{j-k}(s')d_{x_k-x_i}(s)
\label{bad}
\end{equation}
and~\eqref{acp2}, we rewrite the first term of~\eqref{zwa-1} 
as
\begin{align}
\sum_{i=n+1}^{\infty}\sum_{j=1}^\infty z_1^{-i}\A'_{ij}z_2^{j}
&=
\sum_{k=1}^{\infty}z_2^{k}\prod_{j=1}^{2N}
\g^{-1}(z_2,s^{(2j)})
\sum_{i=n+1}^\infty
d_{x_k+i-1}(s)z_1^{-i}\notag\\
&=
\sum_{k=n+1}^{\infty}\left(\frac{z_2}{z_1}\right)^{k}
\prod_{j=1}^{2N}
\g^{-1}(z_1^{-1},s^{(2j-1)})\g^{-1}(z_2,s^{(2j)})
\notag\\
&~~
+\sum_{k=1}^{n}z_2^k\sum_{i=n+1}^{\infty}d_{x_k+i-1}(s)
z_1^{-i}
\prod_{j=1}^{2N}\g^{-1}(z_2,s^{(2j)}).
\label{first}
\end{align}
We need a quite lengthy calculation for the second term of~\eqref{zwa-1}.
At first, by using~\eqref{aij} and~\eqref{bij}, the part 
$\sum_{i=n+1}^\infty z^{-i}b_{ik}$ in the term
can be represented as 
\begin{align}
\sum_{i=n+1}^\infty z_1^{-i}b_{ik}
&=-\sum_{j=1}^n b_{jk}\sum_{i=n+1}^\infty a_{ij}z_1^{-i}\notag\\
&=-\sum_{j=1}^n b_{jk}\sum_{i=n+1}^\infty
\left(\sum_{\ell=0}^{i-n-1}c_{m_j+i-1-\ell}(s)d_{\ell}(s)
+\sum_{a=1}^n c_{m_j+a-1}(s)d_{m_a+i-1}(s)\right)z_1^{-i}.
\label{second1}
\end{align}
In this equation, we notice that the term 
$\sum_{i=n+1}^\infty\sum_{\ell=0}^{i-n-1}c_{m_j+i-1-\ell}(s)d_{\ell}(s)
z_1^{-i}$
becomes
\begin{equation}
\sum_{i=n+1}^\infty\sum_{\ell=0}^{i-n-1}c_{m_j+i-1-\ell}(s)d_{\ell}(s)
z_1^{-i}
=
-\sum_{\ell=0}^{m_j+n-1}c_{m_j+n-1-\ell}(s)z_1^{\ell-n}
\prod_{i=1}^{2N}\g^{-1}(z_1^{-1},s^{(2i-1)})
+z_1^{m_j-1}.
\label{second2}
\end{equation}
The derivation is given as follows. 
By noticing~\eqref{acp1} and~\eqref{acp2},
one gets
\begin{align}
&~\sum_{i=n+1}^\infty\sum_{\ell=0}^{i-n-1}c_{m_j+i-1-\ell}(s)d_{\ell}(s)
z_1^{-i}
\notag\\
&=
-\sum_{i=n+1}^{\infty}\sum_{\ell=i-n}^{m_j+i-1}c_{m_j+i-1-\ell}(s)
d_{\ell}(s)
z_1^{-i}
=-\sum_{i=n+1}^{\infty}\sum_{\ell'=0}^{m_j+n-1}c_{m_j+n-1-\ell'}(s)
d_{\ell'+i-n}(s)z_1^{-i}\notag\\
&=-\sum_{\ell'=0}^{m_j+n-1}c_{m_j+n-1-\ell'}(s) z_1^{\ell'-n}
\left(\sum_{i=n+1}^{\infty}d_{\ell'+i-n}(s)z_1^{-i-\ell'+n}\right)\notag\\
&=-\sum_{\ell'=0}^{m_j+n-1}c_{m_j+n-1-\ell'}(s) z_1^{\ell'-n}
\left(\prod_{i=1}^{2N}\g^{-1}(z_1^{-1},s^{(2i-1)})
-\sum_{i=0}^{\ell'}d_{i}(s)z_1^{-i}\right).
\end{align}
By noting again~\eqref{acp1}, we find
\begin{equation}
\sum_{\ell'=0}^{m_j+n-1}c_{m_j+n-1-\ell'}(s) 
\sum_{i=0}^{\ell'}d_{i}(s)z_1^{\ell'-i-n}
=z_1^{m_j-1}.
\end{equation}
Then, we eventually obtain~\eqref{second2}.
On the other hand, from~\eqref{bad}, 
the other term $\sum_{i=1}^{\infty}\A'_{kj}z_2^{j}$ 
in the second term in~\eqref{zwa-1} is 
\begin{equation}
\sum_{j=1}^{\infty}\A'_{kj}z_2^{j}=
\sum_{b=1}^{k}d_{m_b-m_k}(s)z_2^{b}
\prod_{i=1}^{2N}\g^{-1}(z_2,s^{(2i)}).
\label{second3}
\end{equation}
Thus, from~\eqref{second1},~\eqref{second2}, and~\eqref{second3},
we get  
\begin{align}
&~\frac{1}{b_0}\sum_{k=1}^n\sum_{i=n+1}^\infty z_1^{-i}b_{ik}
\sum_{j=1}^\infty
\A'_{kj}z_2^{j}\notag\\
&=
-\frac{1}{b_0}\sum_{k=1}^n\sum_{j=1}^n b_{jk}z_1^{m_j-1}
\sum_{c=1}^\infty \A'_{kc}z_2^c\notag\\
&~~-\frac{1}{b_0}\sum_{k=1}^{n}\sum_{j=1}^nb_{jk}\sum_{i=n+1}^\infty
\sum_{a=1}^n c_{m_j+a-1}(s)d_{m_a+i-1}(s)z_1^{-i}\sum_{b=1}^k
d_{m_b-m_k}(s)
z_2^b\prod_{i=1}^{2N}\g^{-1}(z_2,s^{(2i)})
\notag\\
&~~+\frac{1}{b_0}\sum_{k=1}^n\sum_{j=1}^n\left(b_{jk}
\sum_{\ell=0}^{m_j+n-1}
c_{m_j+n-1}(s)z_1^{\ell-n}\sum_{b=1}^k d_{m_b-m_k}(s)z_2^b\right)
\prod_{i=1}^{2N}\g^{-1}(z_1^{-1},s^{(2i-1)})
\g^{-1}(z_2,s^{(2i)}).
\label{second4}
\end{align}
Note that the first term in this equation  and the last one 
in~\eqref{zwa-1} cancel each other. We can also find that the 
second term in~\eqref{second4} cancels out the second one 
in~\eqref{first} by the following discussion. 
By using the properties
\begin{align}
&\sum_{k=1}^{n}b_{jk}d_{m_b-m_k}(s)=\tilde{\C}_{(bj)},
\label{det1}\\
&\sum_{j=1}^nc_{m_j+a-1}(s)\tilde{\C}_{(bj)}=\delta_{ab}\det \C
=\delta_{ab}b_0,
\label{det2}
\end{align}
where  $\tilde{\C}_{(bj)}$ is the $bj$ cofactor of the matrix 
$\C$~\eqref{matc},
we have
\begin{align}
&~-\frac{1}{b_0}\sum_{k=1}^{n}\sum_{j=1}^nb_{jk}\sum_{i=n+1}^\infty
\sum_{a=1}^n c_{m_j+a-1}(s)d_{m_a+i-1}(s)z_1^{-i}\sum_{b=1}^k
d_{m_b-m_k}(s)
z_2^b
\prod_{i=1}^{2N}\g^{-1}(z_2,s^{(2i)})
\notag\\
&=-\sum_{a=1}^nz_2^a\sum_{i=n+1}^\infty d_{m_a+i-1}(s)z_1^{-i}
\prod_{i=1}^{2N}\g^{-1}(z_2,s^{(2i)}).
\label{second5}
\end{align}
Furthermore, from relations~\eqref{det1} and~\eqref{det2}, 
we can also rewrite the third term in~\eqref{second4} as 
\begin{align}
&~\frac{1}{b_0}\sum_{k=1}^n\sum_{j=1}^n\left(b_{jk}
\sum_{\ell=0}^{m_j+n-1}c_{m_j+n-1-\ell}(s)z_1^{\ell-n}\sum_{b=1}^k 
d_{m_b-m_k}(s)z_2^b\right)
\prod_{i=1}^{2N}\g^{-1}(z_1^{-1},s^{(2i-1)})
\g^{-1}(z_2,s^{(2i)})
\notag\\
&=
\left(\frac{1}{b_0}\sum_{j=1}^n\sum_{\ell'=0}^{m_j-1}c_{\ell'}(s)
z_1^{m_j-1-\ell'}\sum_{b=1}^{n}\tilde{\C}_{(bj)}z_2^{b}
+\sum_{b=1}^n\left(\frac{z_2}{z_1}\right)^b\right)\notag\\
&~~\times\prod_{i=1}^{2N}\g^{-1}(z_1^{-1},s^{(2i-1)})
\g^{-1}(z_2,s^{(2i)}).
\label{second6}
\end{align}

From~\eqref{second4},~\eqref{second5}, and~\eqref{second6},
we get
\begin{align}
&~\frac{1}{b_0}\sum_{k=1}^n\sum_{i=n+1}^\infty z_1^{-i}b_{ik}
\sum_{j=1}^\infty
\A'_{kj}z_2^{j}\notag\\
&=
-\frac{1}{b_0}\sum_{k=1}^n\sum_{j=1}^n b_{jk}z_1^{m_j-1}
\sum_{c=1}^\infty \A'_{kc}z_2^c
-\sum_{a=1}^nz_2^a\sum_{i=n+1}^\infty d_{m_a+i-1}(s)z_1^{-i}
\prod_{i=1}^{2N}\g^{-1}(z_2,s^{(2i)})
~~\notag\\
&~~
+\left(\frac{1}{b_0}\sum_{j=1}^n\sum_{\ell'=0}^{m_j-1}c_{\ell'}(s)
z_1^{m_j-1-\ell'}\sum_{b=1}^{n}\tilde{\C}_{(bj)}z_2^{b}
+\sum_{b=1}^n\left(\frac{z_2}{z_1}\right)^b\right)\notag\\
&~~\times\prod_{i=1}^{2N}\g^{-1}(z_1^{-1},s^{(2i-1)})
\g^{-1}(z_2,s^{(2i)}).
\label{second}
\end{align}
Hence, from~\eqref{zwa-1},~\eqref{first}, and~\eqref{second},
we finally obtain~\eqref{apB}.

\section*{Acknowledgments}
The work of T.I. is supported by the Core 
Research for Evolutional Science and Technology of Japan Science and 
Technology Agency. The work of T.S. is supported by the Grant-in-Aid 
for Young Scientists (B), the Ministry of Education, Culture, Sports, 
Science and Technology, Japan.

\newpage
%%%%%%%%%%%%%%%%%%%%%%%%%%%%%%%%%
%%% Figure Captions           %%%
%%%%%%%%%%%%%%%%%%%%%%%%%%%%%%%%%
\begin{large}
\noindent
Figure Captions
\end{large}

%%% Fig. 1 %%%%%%%%%%%%%%%%%%%%%%
\vspace{10mm}
\noindent
Fig.~1: 
The random walk interpretation of the weight~\eqref{schurp}.
Note that the $i$th walker starts at the point $-i+1$ and 
returns to the initial point at the end by way of the position
$\lambda^{(n)}_i-i+1$ at time $n$. In this example, the partitions
$\lambda^{(n)}~(n=1,2,3)$ are $\lambda^{(1)}=(4,4,2),~\lambda^{(2)}=(2,1,1),$
and $\lambda^{(3)}=(4,4,3,2,1),$ respectively.

%%% Fig. 2 %%%%%%%%%%%%%%%%%%%%%%
\vspace{10mm}
\noindent
Fig.~2:
The random walk interpretation of the weight~\eqref{pschurp}.
In this case, the $i$th walker starts at the point $1-i$ and
ends at the point $\mu^{(4N)}_i-i+1,$ which is arbitrary but fixed.

%%%%%%%%%%%%%%%%%%%%%%%%%%%%%%%%%
%%% Figures                   %%%
%%%%%%%%%%%%%%%%%%%%%%%%%%%%%%%%%
\newpage

\noindent 
Fig.~1

\begin{picture}(400,250)
\put(0,0){\scalebox{0.7}[0.45]{\includegraphics{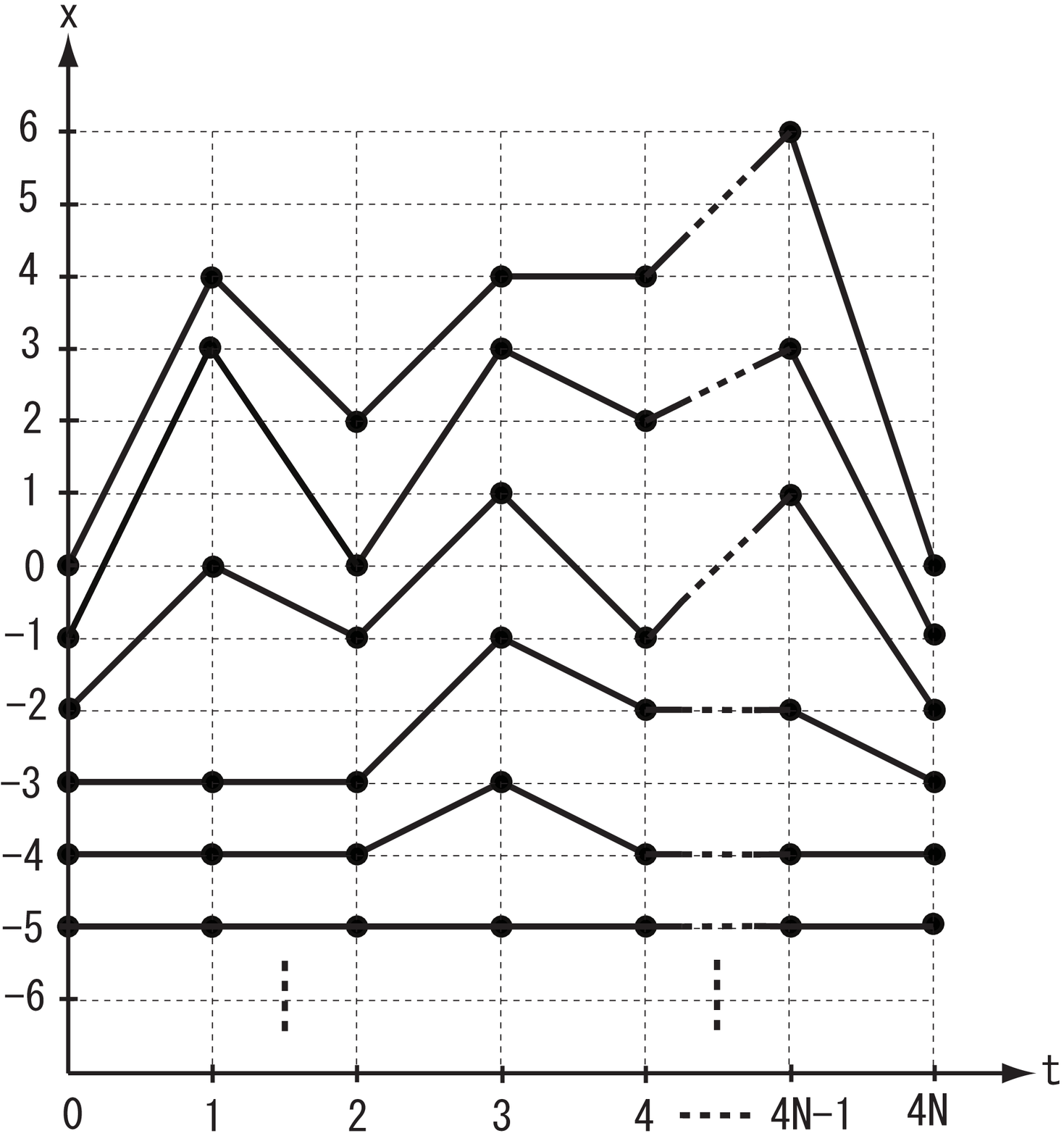}}}
\end{picture}

\vspace{10mm}

\noindent 
Fig.~2

\begin{picture}(400,250)
\put(0,0){\scalebox{0.7}[0.45]{\includegraphics{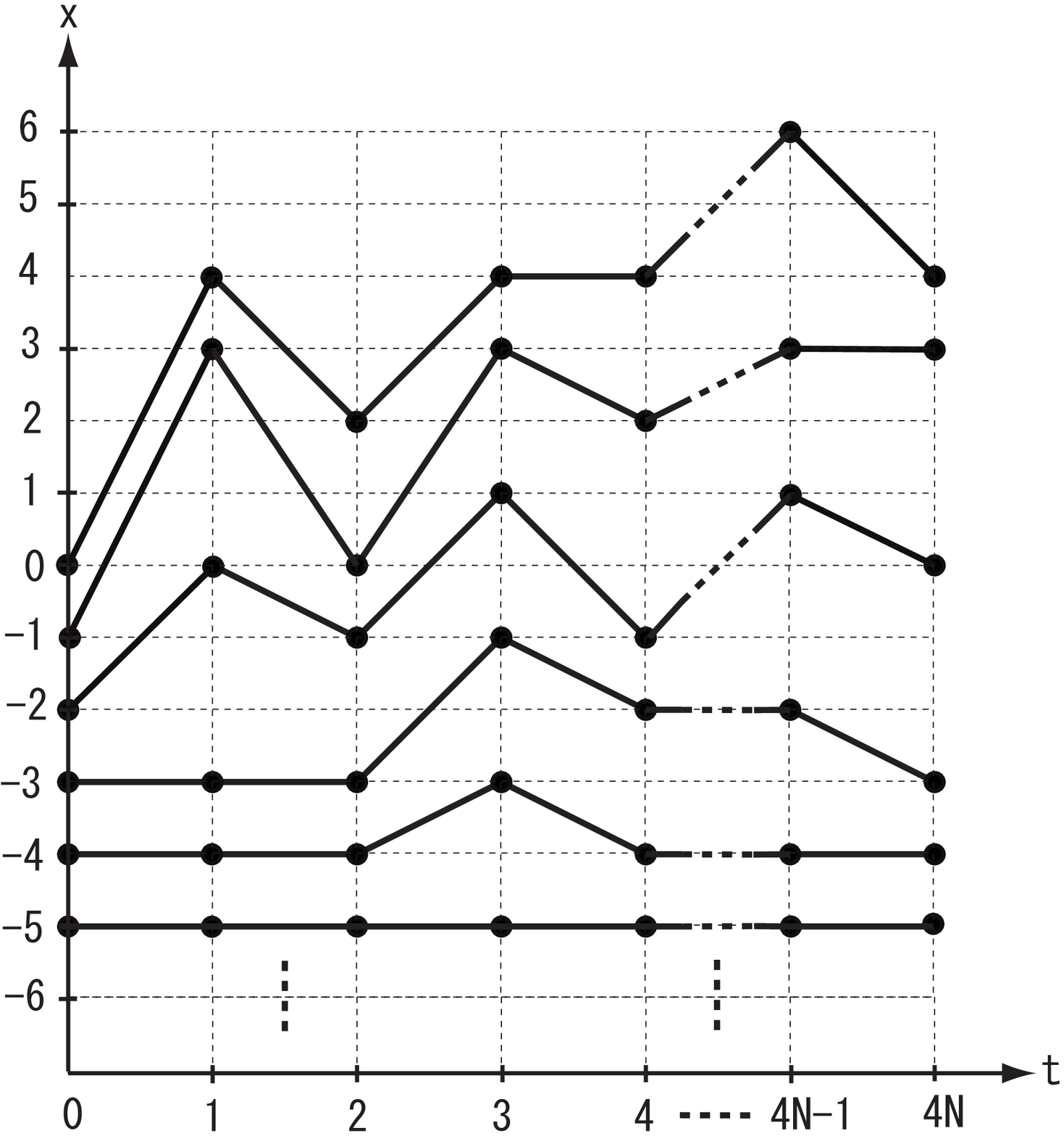}}}
\end{picture}

\end{document}